\documentclass[11pt,a4paper]{article}
\usepackage{subfigure}
\usepackage{xcolor}
\usepackage[T1]{fontenc} 
\usepackage{xspace}
\usepackage{multirow}
\usepackage{hyperref}
\usepackage{slashed}

\usepackage{graphicx,amsmath}
 \usepackage{amssymb}
\usepackage{lineno,color}
\usepackage{url}

\begin{document}
 

\bigskip
\bigskip
    
\begin{center}
{\bf\Large Instanton induced spin-spin correlations }\\
 
 \vspace{1cm}
   
 M.~G.~Ryskin$^{a}$\\
 
 \vspace{0.7cm}
{\small  $^a$ Petersburg Nuclear Physics Institute, NRC Kurchatov Institute, \\Gatchina, St.~Petersburg, 188300, Russia\\
\
}
 \vspace{0.7cm}

 \abstract{\noindent The QCD instanton can be observed at relatively low energies at Nica collider by studying the spin-spin correlations between the incoming proton and the produced $\Lambda$ or $\Sigma$ hyperons (or the $\Lambda$ and
  $\bar\Lambda$ hyperons).}

 \vfill
 
 E-mail: 
  \url{ryskin@thd.pnpi.spb.ru}
 
  \end{center}
 \newpage

 \section{Introduction}
Instantons are non-perturbative  field configurations \cite{BPST} which describe semi-classical transitions between topologically inequivalent vacuum sectors in QCD.
In the semi-classical limit, instantons provide dominant contributions to the path integral and describe quantum tunnelling between different vacuum sectors of the theory \cite{tH,Callan:1976je,Jackiw:1976pf}. Instanton is either directly responsible for generating, or at least contributed to many key aspects of non-perturbative low-energy dynamics of strong interactions~\cite{tHooft:1986ooh,Callan:1977gz,Novikov:1981xi,Shuryak:1982dp,DP,Schafer:1996wv}. These include the role of instantons in the breaking of the $U(1)_A$ symmetry and the spontaneous breakdown of the chiral symmetry, the formation of quark and gluon condensates, $<0|\bar qq|0>$ and $<0|G^a_{\mu\nu}G^a_{\mu\nu}|0>$ and so on.

 Instanton solutions~\cite{BPST} have attracted a lot of interest over the years~\cite{tH,tHooft:1986ooh,Schafer:1996wv,Vainshtein:1981wh,Dorey:2002ik}, but so far instantons have not been observed experimentally in any particle physics settings.

\medskip

The possibility to observe instantons in inelastic proton-proton collisions at hadron colliders was considered in~\cite{BR} and more recently in~\cite{KKS,KMS,12,KKMR,KMRT}. Instanton processes have large production
cross-sections at small centre-of-mass partonic energies~\cite{KKS}, but discovering them at hadron colliders remains challenging~\cite{KMS}.

The problem is that the main signature of the instanton/sphaleron production is the generation of a large number of isotropically distributed mini-jets. That is in inelastic collision we are looking for the high multiplicity event with a large ($S$ close to 1) sphericity. Unfortunately in such a case the background caused by the multiple parton interactions (which also lead to a higher multiplicity and a higher sphericity) in the underlying event is huge.\\

In the present paper we would like to explore another characteristic feature of the QCD instanton. Due to the presence of zero mode for the massless fermion \cite{tH,tHooft:1986ooh} each instanton/sphaleron production is accompanied by the creation of the  light quark-antiquark pairs (one pair for each quark flavour).

Moreover the polarizations of these quarks are fixed.
Instanton($I$) produced the right quarks and the left antiquarks (for each light flavour) while the anti-instanton ($\bar I$, with the opposite topological charge) -- the left quarks and the right antiquarks. \\
\section{Hyperon production}
In terms of Feynman diagrams the 
instanton/sphaleron looks as the non-local vertex with $n_g$ gluon and $2n_f$ fermion legs. Here $n_f$ is the number 
of light quarks with $m_q<1/\rho$
 ($m_q$ is the curent quark mass and $\rho$ is the instanton size).  
 That is 2 (quark + antiquark) legs for each light flavour.
 
 When the instanton was created in the quark-gluon collision the instanton absorbs left-handed quark and emits right-handed quarks
 (anti-instanton absorbs right antiquarks and emits left antiquarks) 
  
$$q_{Li}+g==>I==>n_g\cdot g +q_{Ri}+\sum_f (q_{Rf}+\bar q'_{Lf});\; \; f\neq i,\; 
$$
\begin{equation}
\label{1}
q_{Ri}+g==>\bar I==>n_g\cdot g +q_{Li}+\sum_f (q_{Lf}+\bar q'_{Rf})\quad\quad
n_g\sim 1/\alpha_s(\rho)
\end{equation}

The point is that the strong instanton gluon field rearranges the Dirac basement.
One extra level of light left quark appears while 
the level of right quark goes upstairs to continuum spectra.
\footnote{
This is connected with the $\gamma_5$ anomaly.
In electro-week case where the $\gamma_5$ anomaly is canceled between the quarks and the leptons this leads to the {\em baryon charge non-conservation}.\\
In QCD this is the helicity non-conservation.}\\ 
Note that the instanton not just flips the helicity of the incoming quark $q_i$ (like it happens in the case of the pion or the kaon exchange) but simultaneously create the light quarks of another flavour but with the same helicity.

Thus 
 by studying experimentally the spin-spin correlations, say in
\begin{equation}
\label{2}
p_\uparrow~+~p~\to~\Sigma \ (\mbox{or}\ \Lambda)~+~X
\end{equation}
 process (see Fig.1) we can observe the  effects induced by the instanton production.\\

\begin{figure} [t]
\vspace{-7cm}
\includegraphics[scale=0.61]{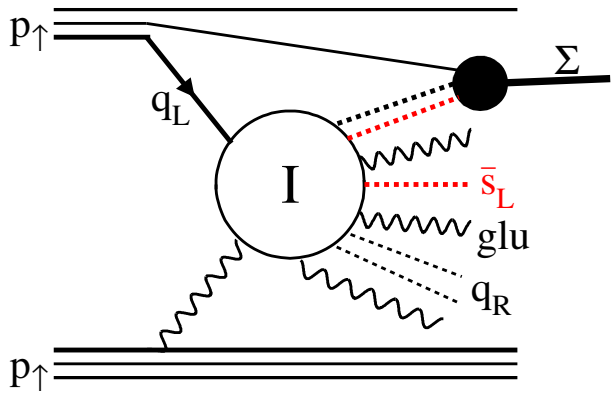}
\begin{center}
\vspace{-1.4cm}
\caption{\small Hyperon production in proton-proton collision mediated by  the instanton/sphaleron.
Dashed lines denote the right-handed quarks and the left-handed antiquarks; strange quark is shown in red.}  
\label{f1}
\end{center}
\end{figure}

First at the quark level the instanton {\em doubles} the incoming polarization. Instead of one left $u_L$-quark 
it produces two right quarks - $u_R$ and $s_R$.
To distinguish the 'left' and 'right' quarks we need 
weak interaction, that is 
weak decay of $\Sigma$ or
 $\Lambda$ hyperons.
 
 Since within the SU(6) quark model the $\Sigma$ hyperon contains the vector $(ud)$ diquark ($\Sigma=s(ud)_V$) there should be a chance to observe in $\Sigma$ the presence of {\em two} right quarks. This is impossible in $\Lambda=s(ud)_S$ case
 where the $(ud)$ diquark is the scalar. On the other hand $\Lambda$ has the advantage -- its polarization (in this model) is equal to the polarization of $s$-quark.\\
 
Another possibility is to produce the $\bar\Lambda\Lambda$ pair and to check that  $\Lambda$ is {\em right}-handed while 
$\bar\Lambda$ is {\em left}-handed on the contrary to the situation where the $\bar ss$ pair was produced perturbatively by the gluon(s). Due to the helicity conservation in quark-gluon vertex here we have to get either two left-handed or two right-handed quarks.\\

To confirm that it was the instanton/sphaleron we could observe a larger than usual (at this energy) multiplicity 
and a  stronger energy ($\sqrt s$) dependence and/or the additional spin-spin correlation when the second beam is polarized and the instanton is created via the quark-quark collision 
 (instanton absorbs two {\em left}-handed quarks only).\\
\subsection{Nica collider}
The corresponding study can be performed at the Nica collider \cite{nika} where the incoming proton beams may be polarized both in the transverse or the longitudinal directions. The energy, $\sqrt s=12-27$  GeV is not large but it will be sufficient to produce the sphaleron of the size $\rho\sim 0.3$ Fm and the mass $M_{inst}\sim 3-5$ GeV.\\
The value $\rho\sim 0.3$ Fm is the typical size of the instantons in the QCD vacuum~\cite{DP-i,ShZ}. Since $1/0.3\mbox{Fm}\simeq 600$ MeV is larger than the current strange quark mass ($m_s\sim 150$ MeV) such an instanton/sphaleron should emit $u,d$ and $s$ quarks plus few ($n_g\sim 1/\alpha_s{\rho}\sim 1-3$) gluons. Each parton has the energy $E\sim 1/\rho=0.6$ GeV. That is to create such a sphaleron we need the energy $M_{inst}\sim 4$ GeV.

It is known that the instanton production cross section steeply increases with $\rho$ (that is with $M_{inst}$ decreasing); see for example the Table 2 of \cite{KKS} where it reaches of about 5 mb already at $M_{inst}=10$ GeV. This is explained mainly by the decreasing of the instanton action $S_{inst}=2\pi/\alpha_s$ (which suppresses as $e^{-S_{inst}}$ the instanton contribution) due to the growth of $\alpha_s(\rho)$.   

Unfortunately at this low scale the precise behavior of QCD coupling $\alpha_s(\rho)$ is not known. Therefore instead of the straightforward calculation we use the simple geometrical estimate
 $\sigma\sim \pi\rho^2\sim 3$ mb. However besides the probability to create the appropriate sphaleron we have to account for the small probability to form the hyperon in the final state. So we would expect the instanton contribution to the hyperon production to be of the order of few $\mu$b.\\
 This is not too small value bearing in mind the Nica luminosity up to $10^{32}$ cm$^{-2}$/sec.  The 1 $\mu$b cross section is comparable with that  observed experimentally in the proton fragmentation region. Thus we hope that it will be possible to observe the instanton induced spin-spin correlations between the hyperons and the polarized incoming proton at the Nica collider.
  \section*{Acknowledgments}
  Author thanks V.A. Khoze for reading the manuscript.

Eur. Phys. J. {\bf C 83} (2023) no.1, 35;  
arXiv:2208.14089 [hep-ph].

\bibitem{Shuryak:2003xz}
E.~Shuryak and I.~Zahed,
Phys. Rev. D \textbf{68} (2003), 034001,
arXiv:hep-ph/0302231 [hep-ph].

\bibitem{nika}
O. Brovko et al.,
JPS Conf.Proc. 35 (2021) 011003,
 Contribution to: STORI'17;\\
    NICA Collaboration • Alexander D. Kovalenko et al.
        PoS SPIN2018 (2019) 007 • Contribution to:
 SPIN 2018, 007.
\bibitem{DP-i} D. Diakonov and V. Yu. Petrov,
 Nucl. Phys. {\bf B 245} (1984) 259.

\bibitem{ShZ}     Edward Shuryak and Ismail Zahed, 
 e-Print: 2102.00256 [hep-ph]


\thebibliography{}

\bibitem{BPST} 
  A.~A.~Belavin, A.~M.~Polyakov, A.~S.~Schwartz and Y.~S.~Tyupkin,
  Phys.\ Lett.\  {\bf 59B} (1975) 85.
    
\bibitem{tH} 
  G.~'t Hooft,
  Phys.\ Rev.\ D {\bf 14} (1976) 343,
   Erratum: [Phys.\ Rev.\ D {\bf 18} (1978) 2199].
\bibitem{Callan:1976je}
  C.~G.~Callan, Jr., R.~F.~Dashen and D.~J.~Gross,
  Phys.\ Lett.\  {\bf 63B} (1976) 334.
  
\bibitem{Jackiw:1976pf}
  R.~Jackiw and C.~Rebbi,
  Phys.\ Rev.\ Lett.\  {\bf 37} (1976) 172.
\bibitem{tHooft:1986ooh}
  G.~'t Hooft,
  Phys.\ Rept.\  {\bf 142} (1986) 357.
  \bibitem{Callan:1977gz}
  C.~G.~Callan, Jr., R.~F.~Dashen and D.~J.~Gross,
  Phys.\ Rev.\ D {\bf 17} (1978) 2717.
  
\bibitem{Novikov:1981xi}
  V.~A.~Novikov, M.~A.~Shifman, A.~I.~Vainshtein and V.~I.~Zakharov,
  Nucl.\ Phys.\ B {\bf 191} (1981) 301.
  
\bibitem{Shuryak:1982dp}
  E.~V.~Shuryak,
  Nucl.\ Phys.\ B {\bf 203} (1982) 116.
    
\bibitem{DP}
  D.~Diakonov and V.~Y.~Petrov,
  Phys.\ Lett.\  {\bf 147B} (1984) 351;
%
 Nucl.\ Phys.\ B {\bf 272} (1986) 457.


\bibitem{Schafer:1996wv}
  T.~Sch{\"a}fer and E.~V.~Shuryak,
  Rev.\ Mod.\ Phys.\  {\bf 70} (1998) 323
  hep-ph/9610451.

%
\bibitem{Vainshtein:1981wh}
  A.~I.~Vainshtein, V.~I.~Zakharov, V.~A.~Novikov and M.~A.~Shifman,
  Sov.\ Phys.\ Usp.\  {\bf 25} (1982) 195
   [Usp.\ Fiz.\ Nauk {\bf 136} (1982) 553].
   

\bibitem{Dorey:2002ik}
  N.~Dorey, T.~J.~Hollowood, V.~V.~Khoze and M.~P.~Mattis,
  Phys.\ Rept.\  {\bf 371} (2002) 231,
  hep-th/0206063.
  
\bibitem{BR}
  I.~I.~Balitsky and M.~G.~Ryskin,
  Phys.\ Atom.\ Nucl.\  {\bf 56}, 1106 (1993)
  [Yad.\ Fiz.\  {\bf 56N8}, 196 (1993)];
  Phys.\ Lett.\ B {\bf 296} (1992) 185.
  
\bibitem{KKS} 
  V.~V.~Khoze, F.~Krauss and M.~Schott,
  JHEP {\bf 2004} (2020) 201,
  arXiv:1911.09726 [hep-ph].

\bibitem{KMS}
  V.~V.~Khoze, D.~L.~Milne and M.~Spannowsky,
  Phys.\ Rev.\ D {\bf 103} (2021) no.1,  014017,
  arXiv:2010.02287 [hep-ph].
  
\bibitem{12}
S.~Amoroso, D.~Kar and M.~Schott,
Eur. Phys. J. C \textbf{81} (2021) no.7, 624
arXiv:2012.09120 [hep-ph].  
  
\bibitem{KKMR}
V.~A.~Khoze, V.~V.~Khoze, D.~L.~Milne and M.~G.~Ryskin,
Phys. Rev. D \textbf{104} (2021) no.5, 054013
arXiv:2104.01861 [hep-ph]. 
  
  \bibitem{KMRT} M. Tasevsky,
V.~A.~Khoze,  D.~L.~Milne and M.~G.~Ryskin 
\end{document}